\documentclass[sigconf,nonacm]{acmart}
\usepackage[nolist, acronym]{glossaries}
\usepackage{multirow}
\usepackage{subcaption}
\usepackage{graphicx}
\usepackage{cleveref}

\AtBeginDocument{%
  }

\makeglossaries
\newacronym{ai}{AI}{Artificial Intelligence}
\newacronym{aoi}{AOI}{Area of Interest}

\newacronym{cv}{CV}{Computer Vision}

\newacronym{et}{ET}{Eye-tracking}
\newacronym{ft}{FT}{Fine-Tuning}
\newacronym{guis}{GUIs}{graphical user interfaces}

\newacronym{infovis}{InfoVis}{Information visualization}
\newacronym{llm}{LLM}{Large Language Model}
\newacronym{llms}{LLMs}{Large Language Models}

\newacronym{ml}{ML}{Machine Learning}

\newacronym{nn}{NN}{Neural Networks}
\newacronym{ocr}{OCR}{Optical Character Recognition}

\newacronym{sota}{SOTA}{State of the Art}
\newacronym{trt}{TRT}{total reading time}

\newacronym{vi}{VI}{Value Iteration}

\begin{document}

\title{A Comparative Study of Scanpath Models in Graph-Based Visualization}
\author{Angela Lopez-Cardona}
\email{angela.lopez.cardona@telefonica.com}
\affiliation{%
  \institution{Telef\'{o}nica Scientific Research}
  \city{Barcelona}
  \country{Spain}
}
\affiliation{%
  \institution{Universitat Politècnica de Catalunya}
  \city{Barcelona}
  \country{Spain}
}

\author{Parvin Emami}
\email{parvin.emami@uni.lu}
\affiliation{%
  \institution{University of Luxembourg}
  \city{Esch-sur-Alzette}
  \country{Luxembourg}}

\author{Sebastian Idesis}
\email{sebastianariel.idesis@telefonica.com}
\affiliation{%
  \institution{Telef\'{o}nica Scientific Research}
  \city{Barcelona}
  \country{Spain}}
  
\author{Saravanakumar Duraisamy}
\email{saravanakumar.duraisamy@uni.lu}
\affiliation{%
  \institution{University of Luxembourg}
  \city{Esch-sur-Alzette}
  \country{Luxembourg}}

\author{Luis A. Leiva}
\email{luis.leiva@uni.lu}
\affiliation{%
  \institution{University of Luxembourg}
  \city{Esch-sur-Alzette}
  \country{Luxembourg}}
  
\author{Ioannis Arapakis}
\email{ioannis.arapakis@telefonica.com}
\affiliation{%
 \institution{Telef\'{o}nica Scientific Research}
 \city{Barcelona}
\country{Spain}}

\renewcommand{\shortauthors}{Lopez-Cardona et al.}


\begin{abstract}
Information Visualization (InfoVis) systems utilize visual representations to enhance data interpretation. Understanding how visual attention is allocated is essential for optimizing interface design. However, collecting Eye-tracking (ET) data presents challenges related to cost, privacy, and scalability. Computational models provide alternatives for predicting gaze patterns, thereby advancing InfoVis research. In our study, we conducted an ET experiment with 40 participants who analyzed graphs while responding to questions of varying complexity within the context of digital forensics. We compared human scanpaths with synthetic ones generated by models such as DeepGaze, UMSS, and Gazeformer. Our research evaluates the accuracy of these models and examines how question complexity and number of nodes influence performance. This work contributes to the development of predictive modeling in visual analytics, offering insights that can enhance the design and effectiveness of InfoVis systems.
\end{abstract}

\maketitle

\noindent \textbf{This is the author's version of this work, provided for personal use only. Redistribution is prohibited. The definitive, published version is available in the ACM ETRA Conference Proceedings, ISBN 979-8-4007-1487-0/2025/05, \url{https://doi.org/10.1145/3715669.3725882}.}

\section{Introduction}
\label{sec:introduction}

\gls{infovis} systems rely on visual elements like charts, graphs, networks, and maps to derive intuitive insights from large datasets, by engaging our visual system's capacity for sensing differences rather than making laborious comparisons between data elements \cite{wang_salchartqa_2024}. In today's information-saturated environment, users must decide what to process, filter, and prioritize. Although using visualizations to address this challenge has been widely discussed~\cite{Falschlunger2016}, InfoVis systems have yet to fully realize their potential in aiding critical decision-making. The effectiveness of \gls{infovis} in responding to critical events hinges on the visual representation of data and the speed at which actionable insights can be derived. We argue that physiological computing holds significant promise for innovation in InfoVis by incorporating human body signals, which can lead to more precise user interface adaptations through closed-loop implicit monitoring~\cite{Fairclough22_neuroadaptive}. This is supported by evidence that cognitive processes are affected by user interface aesthetics, detectable through physiological signals~\cite{hirshfield_this_2011, peck_using_2013, lukanov_using_2016, Haddad24}. Specifically, visual attention allocation in graphic design reflects the perceived importance of design elements~\cite{cartella_trends_2024}. Understanding how users allocate their attention presents a research opportunity~\cite{emami_impact_2024}.  

In the context of visual attention, we can differentiate between bottom-up saliency, which arises from low-level visual features like color, contrast, and motion, capturing attention automatically. In contrast, top-down saliency is shaped by prior knowledge, tasks, and goals, guiding attention based on relevance ~\cite{wang_scanpath_2024}. While bottom-up saliency is driven by stimuli, top-down saliency is cognitive and context-dependent ~\cite{wang_scanpath_2024}. \gls{infovis} mainly involves top-down processing as users engage with visual representations ~\cite{wang_scanpath_2024}. 

To model visual attention, \gls{et} is typically used. However, collecting organic data presents significant challenges, including the reliance on proprietary, high-precision equipment and data privacy concerns \cite{khurana_synthesizing_2023}. Recent studies have addressed the limitations of \gls{et} technology by introducing computational attention models, a promising alternative to predict gaze patterns in response to specific stimuli \cite{deng_eyettention_2023}. In the field of \gls{cv}, predictive models of human attention have demonstrated their utility across various applications, including the optimization of interaction designs and the enhancement of webpage layouts ~\cite{li_uniar_2024}. To train and evaluate these models, scientific datasets containing fixation points generated by human observers exploring images on specific tasks have been developed. These datasets are typically captured using eye-trackers, webcams ~\cite{assens_pathgan_2018} and mouse clicks with techniques like BubbleView \cite{kim_bubbleview_2017}. Popular examples are MASSVIS ~\cite{borkin_beyond_2016}, and UEyes dataset ~\cite{jiang_ueyes_2023} for \gls{infovis} and SALICON ~\cite{jiang_salicon_2015} and CAT2000 ~\cite{borji_cat2000_2015} for natural images. 

Typically, existing solutions predict either visual attention \cite{aydemir_tempsal_2023, chen_learning_2023, fosco_predicting_2020}, also referred to as saliency modeling, or scanpaths, which is also regarded as another form of saliency prediction that accounts for the temporal dimension \cite{martin_scangan360_2022, kummerer_deepgaze_2022, wang_scanpath_2024, sui_scandmm_2023}. Saliency prediction aims to generate heatmaps or probability distributions that emphasize areas of attention within an image ~\cite{cartella_trends_2024}. Contemporary approaches utilize deep neural networks to automatically learn discriminative features, as evidenced by models such as those proposed by ~\citet{chen_learning_2023} and 
~\citet{aydemir_tempsal_2023}. 
~\citet{aydemir_tempsal_2023}.

However, collapsing attention prediction to saliency maps results in the loss of temporal information of attentional deployment~\cite{cartella_trends_2024}. Two scanpaths that cover similar areas of an image, but follow entirely different sequences, can yield identical saliency maps. Nevertheless, the order in which a visual stimulus is explored is important, as it more accurately reflects the prominence of the image elements in relation to the observer~\cite{cartella_trends_2024}. This has driven research into the complementary task of scanpath prediction---the task of forecasting a sequence of fixations over a visual stimulus. Certain models for predicting scanpaths in images have been proposed, with recent efforts placing more emphasis on deep learning solutions (e.g., TPP-Gaze~\cite{damelio_tpp-gaze_2024}). While most models are based on free-viewing human scanpaths, the former ones emerge for \gls{infovis} ~\cite{wang_scanpath_2024}. Specific models are needed since \gls{infovis} stimuli are fundamentally different from natural images: they typically include more text and larger areas with uniform color and minimal to no texture ~\cite{wang_scanpath_2024, matzen_data_2018}.

By modeling eye movement patterns based on visual saliency, we can gain insights into user perception and interaction, reducing the need to involve users early in the design process ~\cite{emami_impact_2024, shi_chartist_2025}. These predictions enable us to measure the impact of visual changes in \gls{infovis}, optimize design components for specific attention levels, and evaluate user awareness of elements over time ~\cite{emami_impact_2024}. Such solutions could be used as an evaluation tool, which visualization designers can use to compare candidate designs in either a qualitative or quantitative manner ~\cite{matzen_data_2018}. Modelling scanpaths on \gls{infovis} can offer insights into the rich spatial and temporal dynamics of human attention~\cite{wang_scanpath_2024}.

In this paper, we conducted a \gls{et} study with 40 participants who analysed graphs of different numbers of nodes, while answering questions of varying complexity in the context of digital forensics.  We investigated how similar these gaze patterns are to synthetic scanpaths produced by various generative models: DeepGaze ~\cite{kummerer_deepgaze_2022}, UMSS ~\cite{wang_scanpath_2024} and Gazeformer ~\cite{mondal_gazeformer_2023}. Our objective is to determine which state-of-the-art generative model can most effectively synthesise realistic scanpaths on graph-based visualizations and how the question and visualisation complexity impacts their accuracy.

\section{Related work}


\acrlong{et} is extensively used in \gls{infovis} as it offers valuable insights into information foraging and decision-making processes, thereby informing research on visual attention
~\cite{borkin_beyond_2016} ~\cite{nguyen_iteractive_2017}.

Considering the different conditions in visual analysis tasks, ~\citet{polatsek_exploring_2018} explored how visual attention and perception influence task-based analysis. By examining the MASSVIS dataset, their study demonstrated that free-viewing saliency models fall short in capturing the nuances of task-driven attention, suggesting that these models require further refinement.
Following that, \citet{wang_salchartqa_2024} created the SalChartQA dataset, a human-annotated Chart Question Answering dataset covering three visualization types: bar plots, line plots, and pie charts. Analyses on SalChartQA demonstrate the strong impact of the question on visual saliency. However, this study does not analyze graph-based visualizations and collects data using BubbleView ~\cite{kim_bubbleview_2017}.


Early research on scanpath prediction has primarily utilized bottom-up saliency maps to predict gaze shifts and positions. Notable models, such as SalTiNet ~\cite{assens_saltinet_2017}, introduced a deep learning framework that innovatively represents saliency maps in three dimensions, termed saliency volumes. This structure captures the temporal aspect of fixations by adding a temporal axis to traditional saliency maps. Scanpaths are generated by sampling fixation points from these saliency volumes, followed by a post-filtering stage. PathGAN \cite{assens_pathgan_2018} advanced this by offering a fully end-to-end solution, employing a generative adversarial network (GAN) to directly produce scanpaths without the need for sampling or post-processing. DeepGaze III \cite{kummerer_deepgaze_2022} improved prediction using saliency maps and previous scanpaths to forecast subsequent fixations, establishing itself as the state-of-the-art for generating free-viewing human scanpaths. Later, \citet{jiang2023ueyes} introduced DeepGaze++ based on DeepGaze III with modifications and fine-tuned it on the UEyes dataset. \citet{emami_impact_2024} optimized the parameters of DeepGaze++, such as masking radius, IOR mechanisms and input image size for user interfaces, relevant to \gls{infovis}. So we consider this version of DeepGaze++ as the state-of-the-art model for scanpath prediction with these parameters.


Recently, there has been a growing interest in visual attention models for \gls{infovis}, as demonstrated by \citet{matzen_data_2018} who examined the limitations of existing saliency models on the MASSVIS dataset. Their findings motivated the development of the Data Visualization Saliency (DVS) model to better address these shortcomings. However, scanpath prediction remains underexplored, with UMSS \cite{wang_scanpath_2024} emerging as the first notable contribution and serving as the state-of-the-art model in this area. Moreover, current methods typically generate visual attention maps without considering user questions or queries \cite{shi_chartist_2025}. The investigation of goal-directed attention---such as that required for visual search tasks---remains in its early stages \cite{cartella_trends_2024}. In \gls{infovis}, research on question-driven saliency and scanpath prediction is especially recent.

For saliency, VisSalFormer \cite{wang_salchartqa_2024} remains the only method available for predicting question-driven saliency on information visualizations. In terms of scanpaths, Chartist \cite{shi_chartist_2025} is the only available model; it simulates how users move their eyes to forage information on a chart---handling tasks such as value retrieval, filtering, and identifying extremes. Although Chartist was developed around the same time as this work, its code was not available, so a direct comparison was ommitted from our study. Moreover, since it is trained in a controlled environment rather than using human eye-tracking data, it opens a promising direction for future research.


 In the context of visual question answering, models like Chen et al. \cite{chen_predicting_2021} combine visual-question-answering frameworks with a ConvLSTM module to predict fixation position distributions and durations based on questions related to natural images. Recently, Gazeformer \cite{mondal_gazeformer_2023} proposed using a Transformer model to regress fixation coordinates within a scanpath focused on object searching, conditioned on the embedding of the object's name from a pre-trained language model. For all these reasons, we considered Gazeformer to be the state-of-the-art model for scanpath prediction in goal-directed attention. 

Although these models are useful, they often depend on task-specific configurations that limit their generalizability. UniAR \cite{li_uniar_2024} overcomes this by predicting scanpaths and saliency while also incorporating user input preferences across tasks. While promising, its closed-source nature prevented us from including it in our work.

\section{Methodology}
\label{sec:method}

This section describes the methodology employed in this study, detailing the data collection process, the generative models used, and the evaluation metrics. The Data Collection subsection outlines participant recruitment \Cref{subsec:particpants}, the stimuli used \Cref{subsec:stimuli}, task procedure\Cref{subsec:exp_setup}, and the experimental apparatus \Cref{subsec:data_adquisition}. The Generative Models subsection introduces the models applied in the study \Cref{subsec:generative} followed by a description of the metrics used for evaluation \Cref{subsec:metrics}.

\subsection{Data Collection}
\label{subsec:data_collection}
The data was collected with the purpose of exploring the impact of different \gls{infovis} adaptations in the context of digital forensics. However, the focus of this work is to investigate the impact of question complexity and number of nodes, and not the nature of the adaptation.

\subsubsection{Participants}
\label{subsec:particpants}

We recruited 40 participants with a mean age of 27.2 years (SD=4.9) through mailing lists and university portal advertisements. Inclusion criteria required participants to be aged 18 to 65 years, have no history of neurological disorders and have normal or corrected-to-normal vision. Participants also needed to understand and follow the experiment instructions and provide written informed consent before participating.

\subsubsection{Stimuli}
\label{subsec:stimuli}

We consider undirected graphs, to convey information about criminal records, such as numbers of arrests or incarceration states. This is the most common visualization type in this domain~\cite{Osborne11, Bohm21}. In this environment, nodes represent people, while the edges indicate their relationships. We followed a within-subjects design, where all participants were exposed to all experimental conditions but in randomized order.

We consider two levels of visual complexity: graphs with three (3) or six (6) nodes. Each graph was presented as one of three different configurations:
\begin{description}
    \item[Baseline:] Graphs without any visual adaptation (see example in Figure~\ref{fig:teaser:baseline}).
    \item[Partial adaptation:] Graphs with only one visual adaptation applied (see examples in Figure~\ref{fig:teaser:edge-color}-Figure~\ref{fig:teaser:node-color}).
    \item[Full adaptation:] Graphs with all adaptations applied simultaneously (see example in Figure~\ref{fig:teaser:all}).
\end{description}

We consider five possible visual adaptations,
based on principles of effective information visualization~\cite{Gelman13, Shen21, Knoblich24} and popular guidelines,\footnote{\url{https://datavizcatalogue.com}} where for example color is used for categorization and size is used to convey quantitative information:
\begin{figure*}[!ht]
    \centering
    \def\w{0.13\linewidth}
    \subfloat[Baseline\label{fig:teaser:baseline}]{\includegraphics[width=\w]{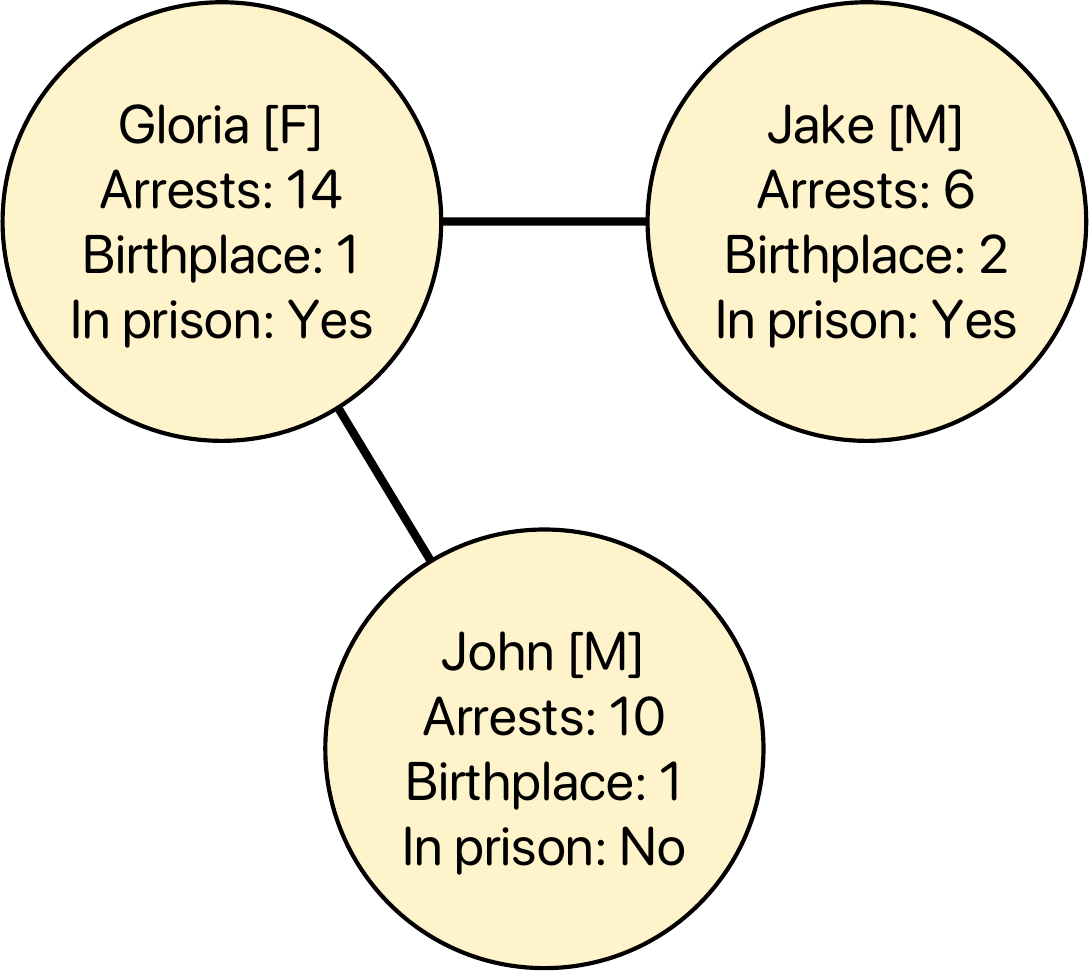}}\hfill
    \subfloat[Edge color\label{fig:teaser:edge-color}]{\includegraphics[width=\w]{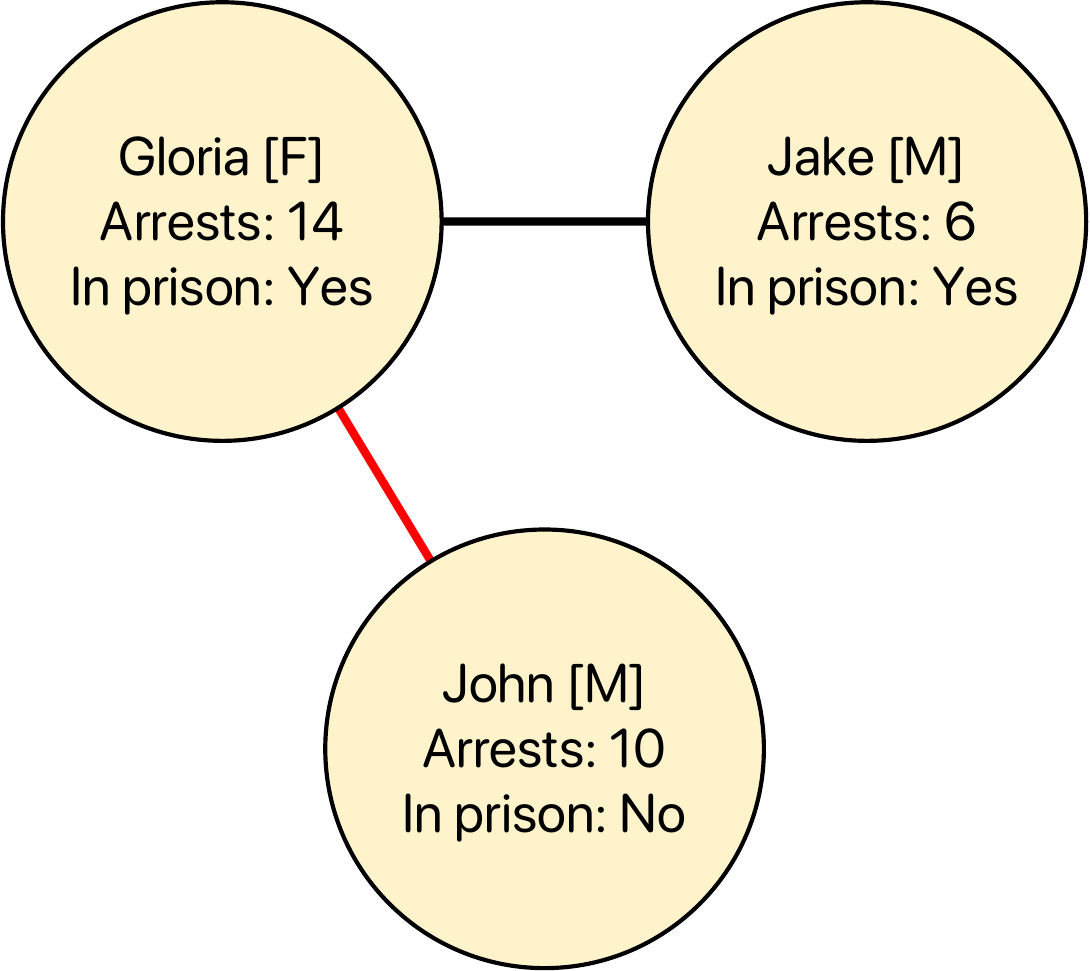}}\hfill
    \subfloat[Edge thickness\label{fig:teaser:edge-thick}]{\includegraphics[width=\w]{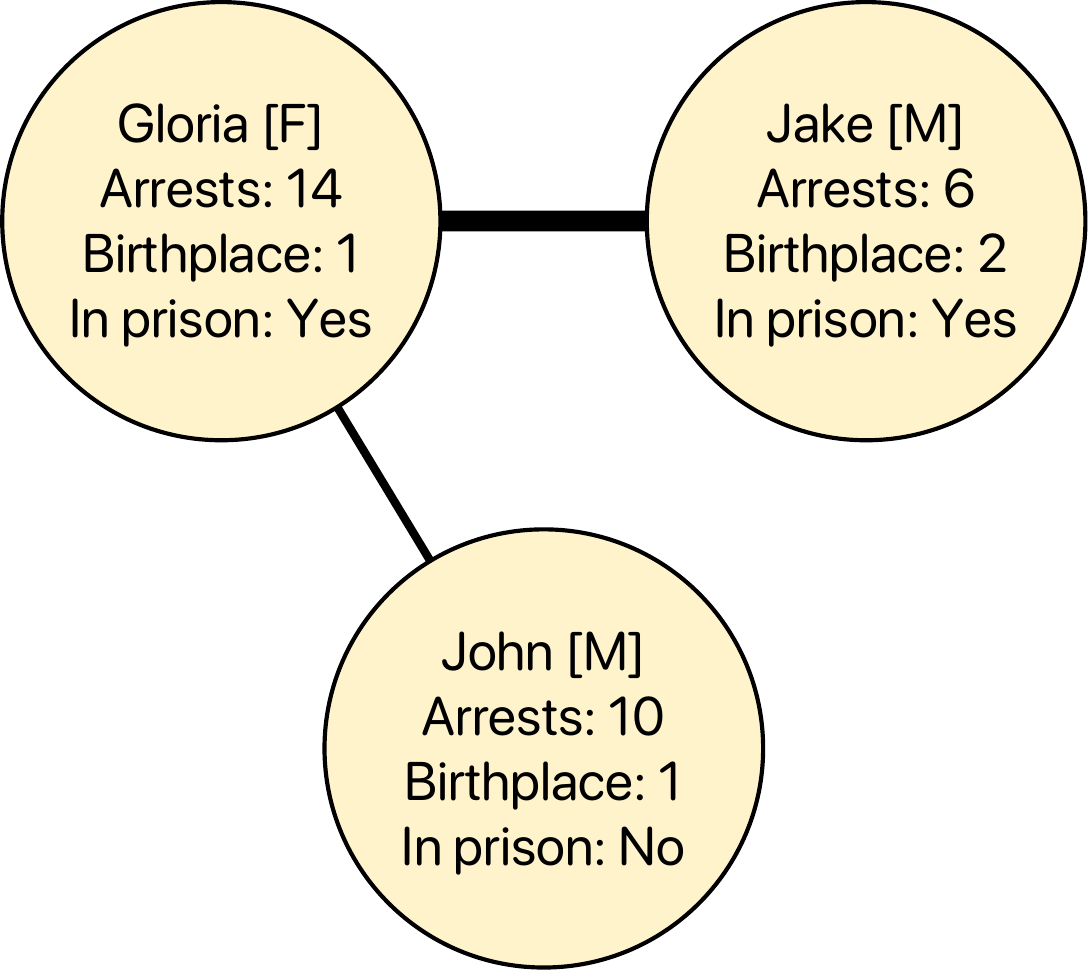}}\hfill
    \subfloat[Node size\label{fig:teaser:node-size}]{\includegraphics[width=\w]{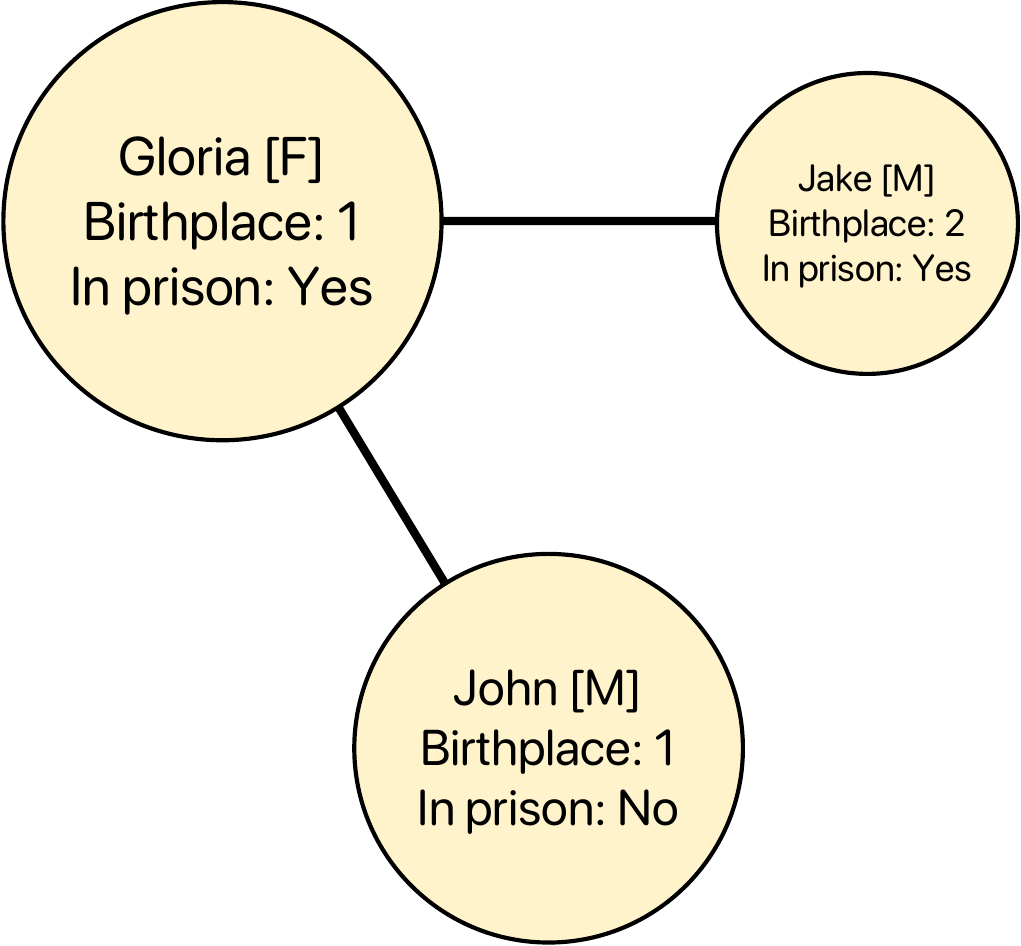}}\hfill
    \subfloat[Node shape\label{fig:teaser:node-shape}]{\includegraphics[width=\w]{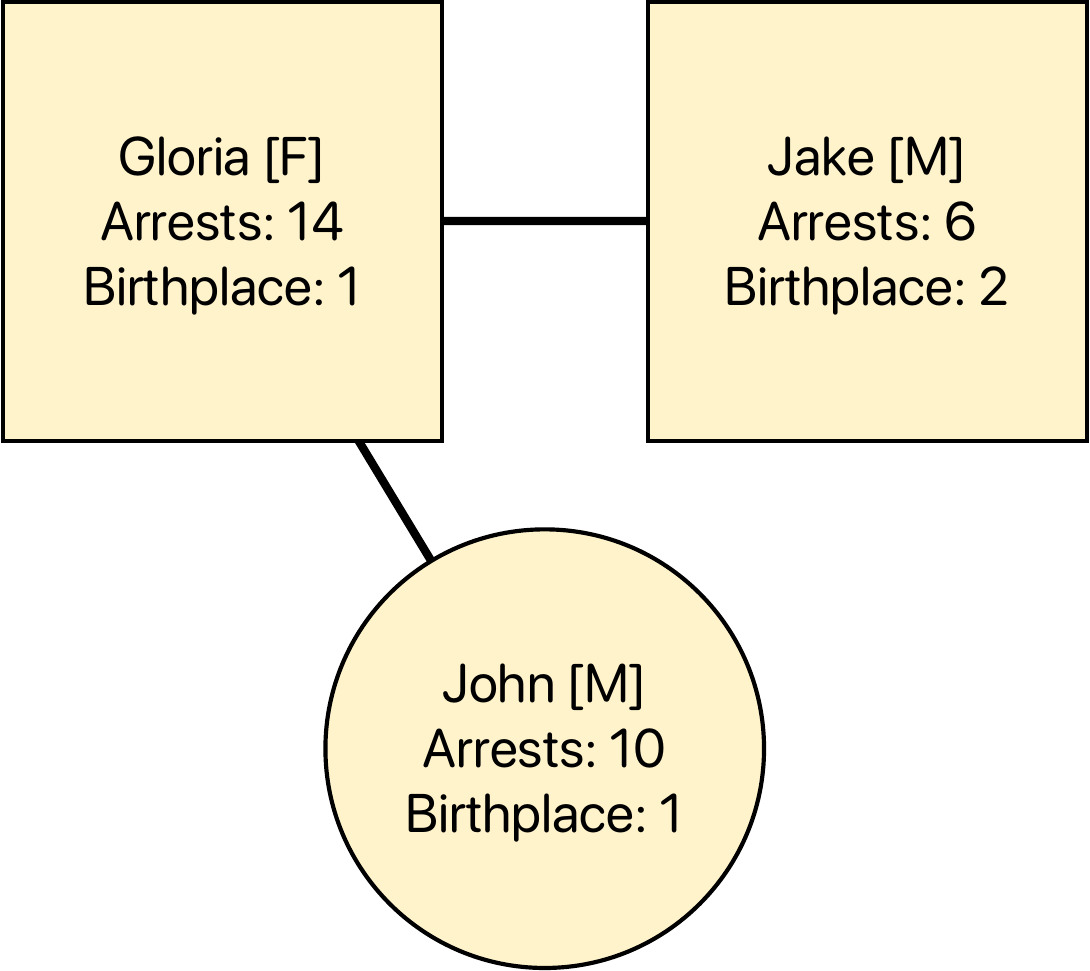}}\hfill
    \subfloat[Node color\label{fig:teaser:node-color}]{\includegraphics[width=\w]{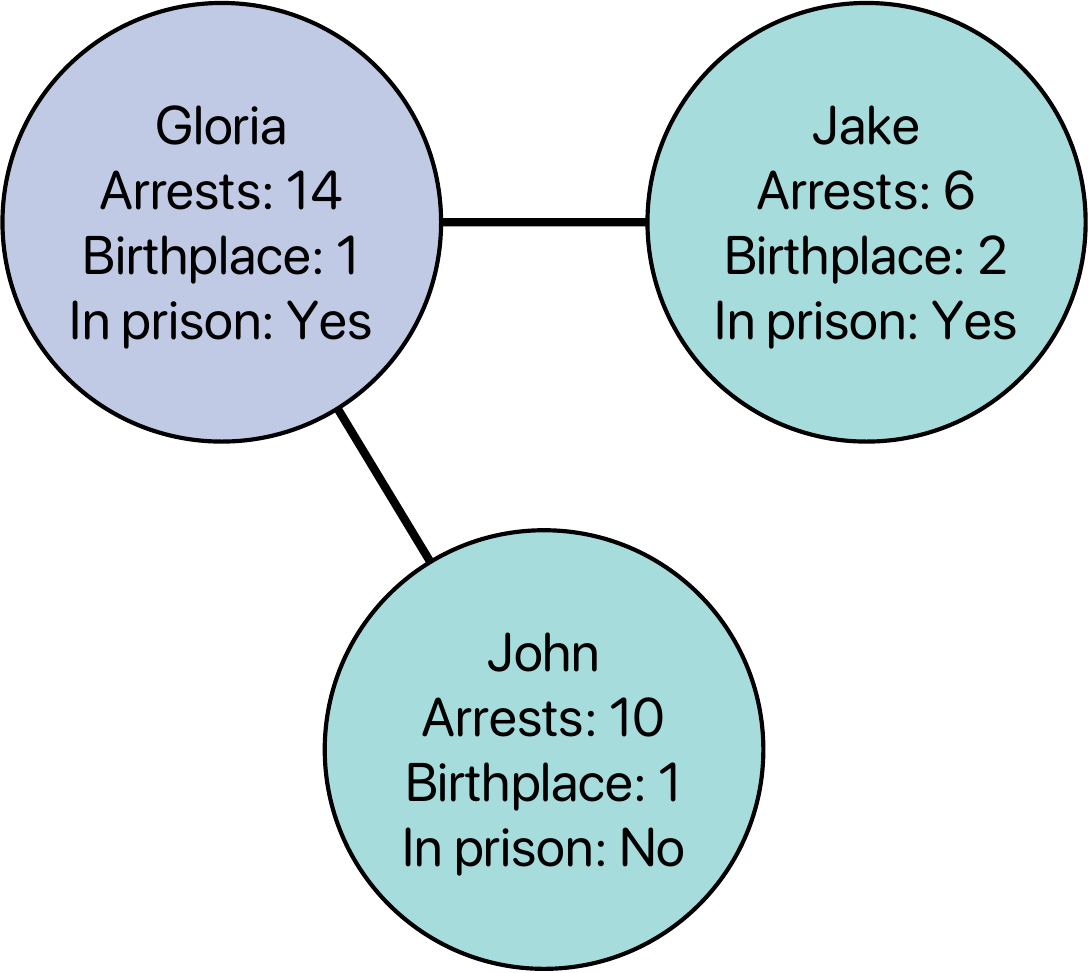}}\hfill
    \subfloat[Full adaptation\label{fig:teaser:all}]{\includegraphics[width=0.12\linewidth]{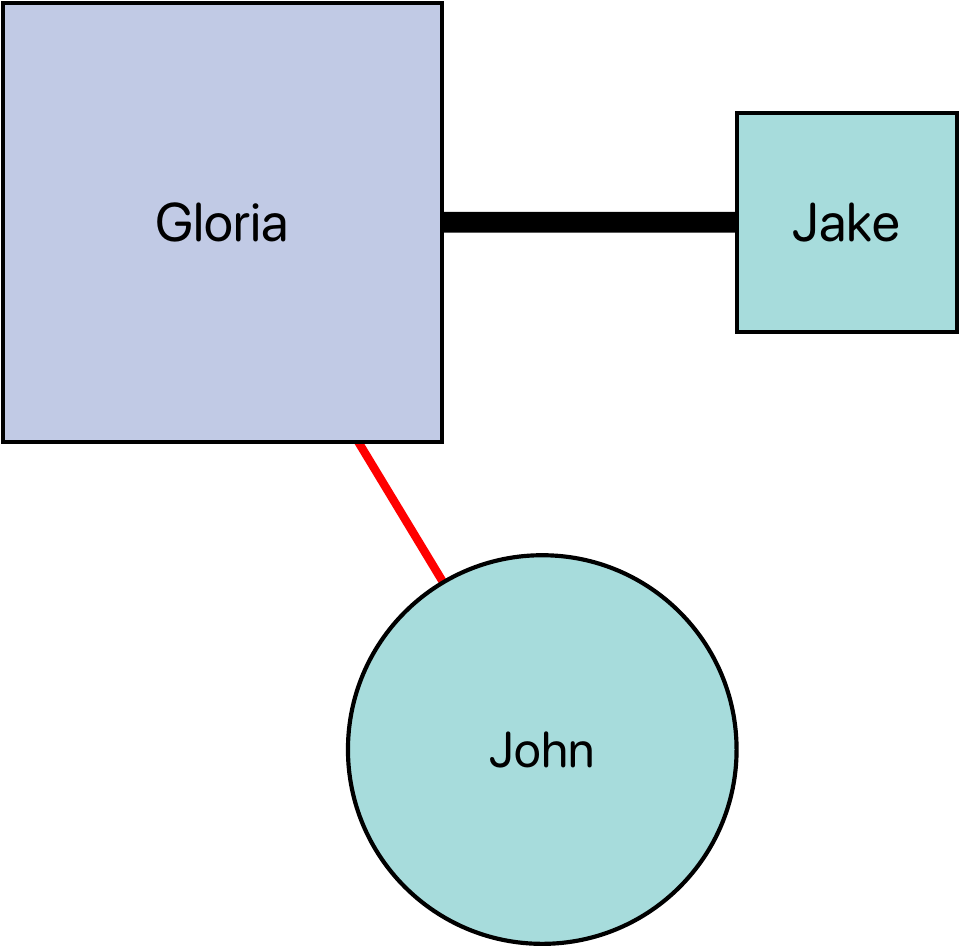}}
 
    \caption{Examples of graph adaptations, from baseline (a)  to fully adapted (g) versions. Figures \protect b-\protect f represent the five possible partial adaptations.}
    \label{fig:teaser}
\end{figure*}    
\begin{description}
    \item[Edge color:] Indicates whether individuals share matching birthplaces (Figure~\ref{fig:teaser:edge-color}).
    \item[Edge thickness:] Represents the strength of the connection between individuals (Figure~\ref{fig:teaser:edge-thick}).
    \item[Node size:] Represents the number of arrests made by an individual (Figure~\ref{fig:teaser:node-size}).
    \item[Node shape:] Distinguishes whether the person is currently in jail (Figure~\ref{fig:teaser:node-shape}).
    \item[Node color:] Indicates the gender of the individual (Figure~\ref{fig:teaser:node-color}).
\end{description}

We prepared a set of questions to assess the participants' ability to interpret the graphs under varying levels of complexity. The questions designed to be answered based on the information conveyed in the graph and were categorized according to their difficulty:
\begin{description}
    \item[Easy questions:] These involved simple queries 
        that required a straightforward interpretation of the graph 
        (e.g., How many people are in jail?).
    \item[Hard questions:] These required a more in-depth analysis 
        and a higher level of cognitive processing 
        (e.g., How many people with the same birthplace are connected to people with more than two arrests?).
\end{description}

Our independent variables are the numbers of nodes (three and six) and question difficulty (two factors: easy and hard).

\subsubsection{Task and Procedure}
\label{subsec:exp_setup}
\phantom{X}
\\

\begin{figure}[!ht]  
\centering  
\includegraphics[width=0.93\linewidth]{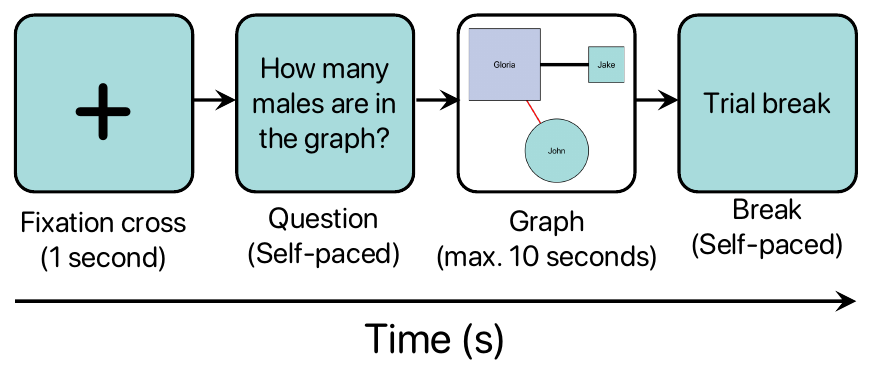}  
\captionof{figure}{ 
Each trial followed a structured sequence: fixation cross, question screen, graph screen, and resting period. This procedure was repeated for all 120 trials.}  
\label{fig:Trial}  
\end{figure}

Participants completed 15 practice trials to familiarize themselves with the task. The experiment was conducted in a dimly lit room to minimize distractions and optimize eye-tracking data quality. The main experiment consisted of 120 trials per participant (30 for each of our analysed categories).
Each trial began with a 1-second fixation cross, followed by a self-paced question-reading phase (Figure~\ref{fig:Trial}). Participants then viewed a graph and had 10 seconds to respond before the system advanced to the next trial automatically. Participants provided single-digit responses as quickly and accurately as possible or pressed ‘n’ if they could not answer. A self-paced resting period followed each trial before proceeding. This cycle repeated for 120 trials, with the experiment lasting about 30 minutes and all trials randomized.

\subsubsection{Apparatus}
\label{subsec:data_adquisition}
Eye movement data was collected using the GP3 HD Eye Tracker, sampling at 60 Hz in accordance with recommended guidelines \cite{leiva2024modeling}. This equipment provides high spatial and temporal resolution, allowing for precise tracking of rapid eye movements \cite{cuve2022validation}. A five-point calibration was conducted for each participant before the study, and it was repeated until minimal error was achieved. Throughout the experiment, various eye movements—including fixations, saccades, and blinks—were recorded. The data collected included timestamps, gaze coordinates (x and y), and eye validity codes, with continuous monitoring of data quality through real-time feedback \cite{duchowski2017eye}.
\subsection{Generative models}
\label{subsec:generative}

\subsubsection{UMSS}
\label{subsec:umss}

Proposed in ~\cite{wang_scanpath_2024}, it involves two main components. The first one produces multi-duration saliency maps for visualization elements and is specifically trained to capture how attention shifts over time based on the visual elements and their relevance at different viewing stages—specifically at 0.5, 3, and 5 seconds. It is based on the model MultiDuration Saliency Excited Model (MD-SEM)\cite{fosco_how_2020} that was proposed for natural images. By calculating the Element Fixation Density (EFD), researchers created the MASSVIS Multi-Duration Element Fixation Density (MASSVIS-MDEFD) using the MASSVIS dataset.
This dataset was subsequently utilized to fine-tune MD-SEM, resulting in the development of their Multi-Duration Element Attention Model (MD-EAM). The second component involves Probabilistic Scanpath Sampling using these saliency maps. UMSS probabilistically samples gaze locations to predict the sequence of fixations.

We have successfully replicated their training process and computed both saliency maps and scanpath predictions over a duration of five seconds, consistent with their original implementation. In our images, the stimuli (nodes) are typically positioned at the top or bottom rather than in the center, which makes the initial scanpath sampling algorithm ineffective. To customize the model for our data, we introduce a weighting factor that integrates the saliency prediction with its Gaussian mask during the scanpath sampling process.

\subsubsection{DeepGaze++}
\label{subsec:deepgaze}
DeepGaze III~\cite{kummerer_deepgaze_2022} generates a probabilistic density map for the next fixation based on input images and previously observed fixation points. This approach can produce clusters of closely spaced points, potentially causing stagnation in those regions. To address this issue, DeepGaze++ ~\cite{jiang_ueyes_2023} iteratively selects the position with the highest probability from the density map and applies a custom inhibition-of-return (IOR) decay to suppress that location in the saliency map. However, the default IOR decay mechanism is effective only for a limited number of fixations. Therefore, in other research, they adopt the IOR decay introduced in~\cite{emami_impact_2024}, which extends coverage beyond 12 fixation points. Furthermore, they integrated the design parameters for this model, which was introduced in the same paper. Despite these adjustments, DeepGaze++ remains a state-of-the-art scanpath model and is thus employed in their investigation. In this paper, we used this model with the IOR method introduced in~\cite{emami_impact_2024}, along with the optimal parameters from the same paper.

\subsubsection{Gazeformer}

\citet{mondal_gazeformer_2023} proposed a transformer-based model for goal-directed human gaze prediction that encodes target objects using a language model, allowing it to generalize to unseen categories (ZeroGaze). The architecture consists of a ResNet-50 \cite{he_resnet_2016} backbone for visual feature extraction and RoBERTa \cite{liu_roberta_2019} for target encoding, which are combined into a shared visual-semantic space. A transformer encoder contextualizes the scene, while a transformer decoder predicts fixation sequences in parallel using learnable fixation queries. Fixation locations and durations are directly regressed using a Gaussian distribution-based approach. Training is formulated as a sequence modeling task with a multi-task loss, including L1 regression loss for fixation coordinates and durations, and a negative log-likelihood loss for fixation validity. The model is trained on the COCO-Search18 dataset ~\cite{chen_coco-search18_2021}. 

In our work, we use their available implementation and trained checkpoint. We encode our questions as the task using the language model. Since the model was originally trained to predict up to seven fixations per stimulus, we likewise restrict our analysis to a maximum of seven fixations.

\begin{table*}[htb]
\centering
\begin{tabular}{llcccc}
\toprule
Question & Graph & \textbf{DTW ($\downarrow$)} & \textbf{Eyenalysis ($\downarrow$)} & \textbf{Determinism ($\uparrow$)} & \textbf{Laminarity ($\uparrow$)} \\
\midrule
hard & 6 nodes & 4.3676 $\pm$ 0.4010 & \textbf{0.0325 $\pm$ 0.0071} & \textbf{1.6162 $\pm$ 1.4088} & \textbf{13.3448 $\pm$ 5.9174} \\
easy & 6 nodes & \underline{4.3009 $\pm$ 0.5005} & \underline{0.0352 $\pm$ 0.0100} & \underline{1.0073 $\pm$ 0.9065} & \underline{8.7668 $\pm$ 4.0127} \\
hard & 3 nodes & 4.3414 $\pm$ 0.4041 & 0.0373 $\pm$ 0.0077 & 0.8511 $\pm$ 0.7929 & 8.5293 $\pm$ 4.7948 \\
easy & 3 nodes & \textbf{4.1614 $\pm$ 0.3345} & 0.0454 $\pm$ 0.0143 & 0.6084 $\pm$ 0.7561 & 7.0486 $\pm$ 4.8389 \\
\bottomrule
\end{tabular}

\caption{ Metrics for the UMSS model (mean $\&$ standard deviation) predicting 12 fixations, evaluated by graph complexity and question difficulty.}
\label{tab:results_umss}
\end{table*}

\begin{table*}[htb]
\centering
\begin{tabular}{llcccc}
\toprule
Question & Graph & \textbf{DTW ($\downarrow$)} & \textbf{Eyenalysis ($\downarrow$)} & \textbf{Determinism ($\uparrow$)} & \textbf{Laminarity ($\uparrow$)} \\
\midrule
hard & 6 nodes & 4.7764 $\pm$0.7233 & \textbf{0.0253 $\pm$0.0078} & \textbf{5.7138 $\pm$2.8525} & \textbf{22.3863 $\pm$7.9269} \\
easy & 6 nodes & \underline{4.3233 $\pm$0.8756} & \underline{0.0315 $\pm$0.0098} & \underline{4.7295 $\pm$2.8249} & \underline{19.5152 $\pm$6.1463} \\
hard & 3 nodes & 4.4060 $\pm$0.6973 & 0.0348 $\pm$0.0085 & 3.8870 $\pm$2.2667 & 13.9008 $\pm$4.7745 \\
easy & 3 nodes & \textbf{3.6790 $\pm$0.4919} & 0.0327 $\pm$0.0084 & 3.6320 $\pm$2.3660 & 17.8022 $\pm$4.7747 \\
\bottomrule
\end{tabular}
\caption{ Metrics for the DeepGaze++ model (mean $\&$ standard deviation) predicting 12 fixations, evaluated by graph complexity and question difficulty.}
\label{tab:results_deepgaze}
\end{table*}

\begin{table*}[htb]
\centering
\begin{tabular}{llcccc}
\toprule
Question & Graph & \textbf{DTW ($\downarrow$)} & \textbf{Eyenalysis ($\downarrow$)} & \textbf{Determinism ($\uparrow$)} & \textbf{Laminarity ($\uparrow$)} \\
\midrule
hard & 6 nodes & 4.9047 $\pm$0.9595 & \underline{0.0392 $\pm$0.0121} & 0.3598 $\pm$0.9258 & \underline{6.4004 $\pm$4.3728} \\
easy & 6 nodes & \underline{3.3858 $\pm$1.2919} & \textbf{0.0379 $\pm$0.0161} & \textbf{1.2398 $\pm$2.2932} & \textbf{7.7372 $\pm$6.3383} \\
hard & 3 nodes & 3.5733 $\pm$1.1852 & 0.0414 $\pm$0.0132 & 0.7707 $\pm$1.8329 & 4.3177 $\pm$3.7336 \\
easy & 3 nodes & \textbf{2.6133 $\pm$0.7640} & 0.0421 $\pm$0.0167 & \underline{0.9666 $\pm$2.3136} & 5.5023 $\pm$5.1348 \\
\bottomrule
\end{tabular}
\caption{ Metrics for the Gazeformer model (mean $\&$ standard deviation) predicting 7 fixations, evaluated by graph complexity and question difficulty.}
\label{tab:results_gazeformer}
\end{table*}

\begin{figure*}[htb]
    \centering
    \def\w{0.25\linewidth}
    \subfloat[Ground-truth\label{fig:scanpath_baseline}]{\includegraphics[width=0.227\linewidth]{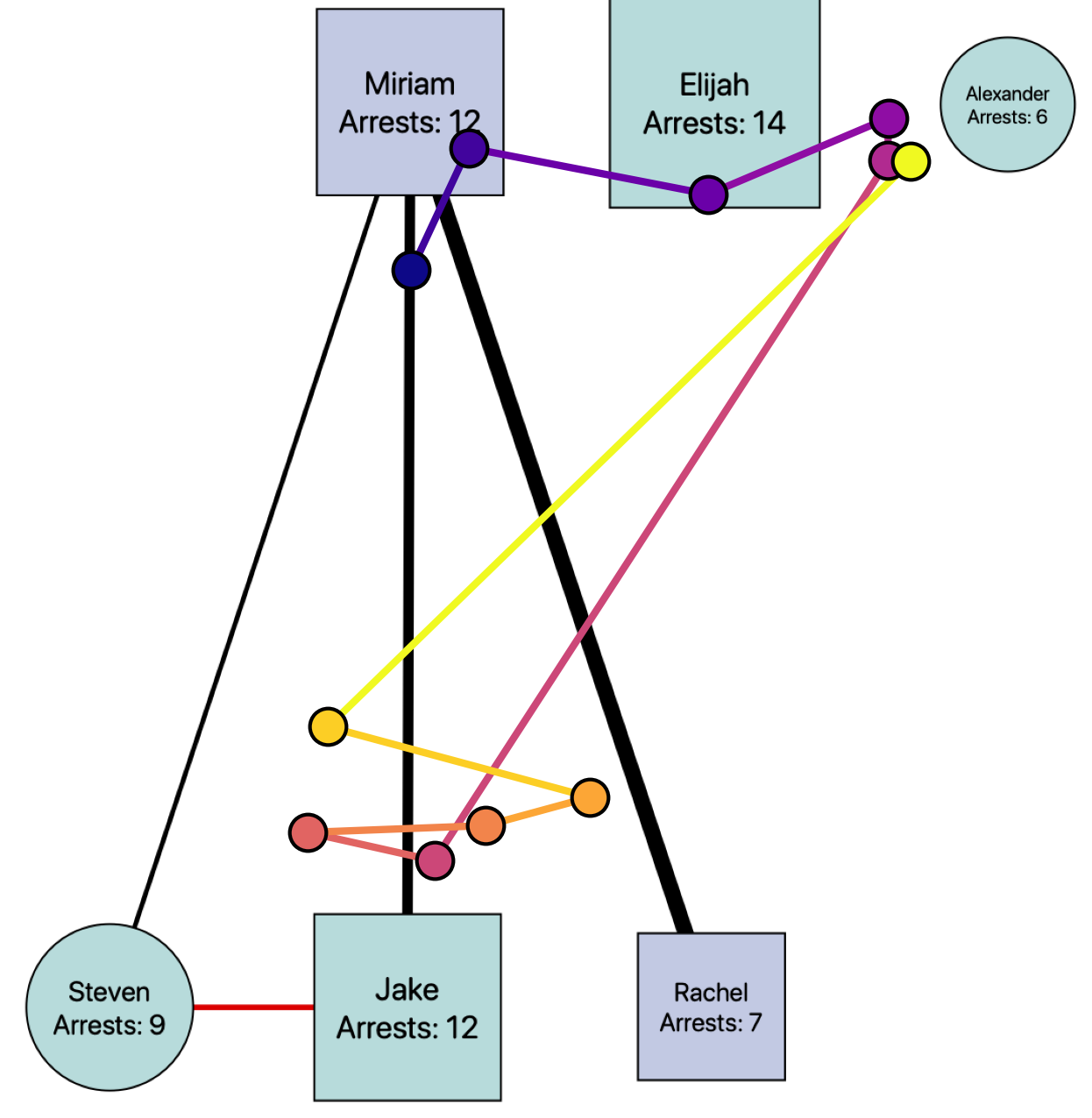}}\hfill
    \subfloat[UMSS\label{fig:scanpath_umss}]{\includegraphics[width=0.235\linewidth]{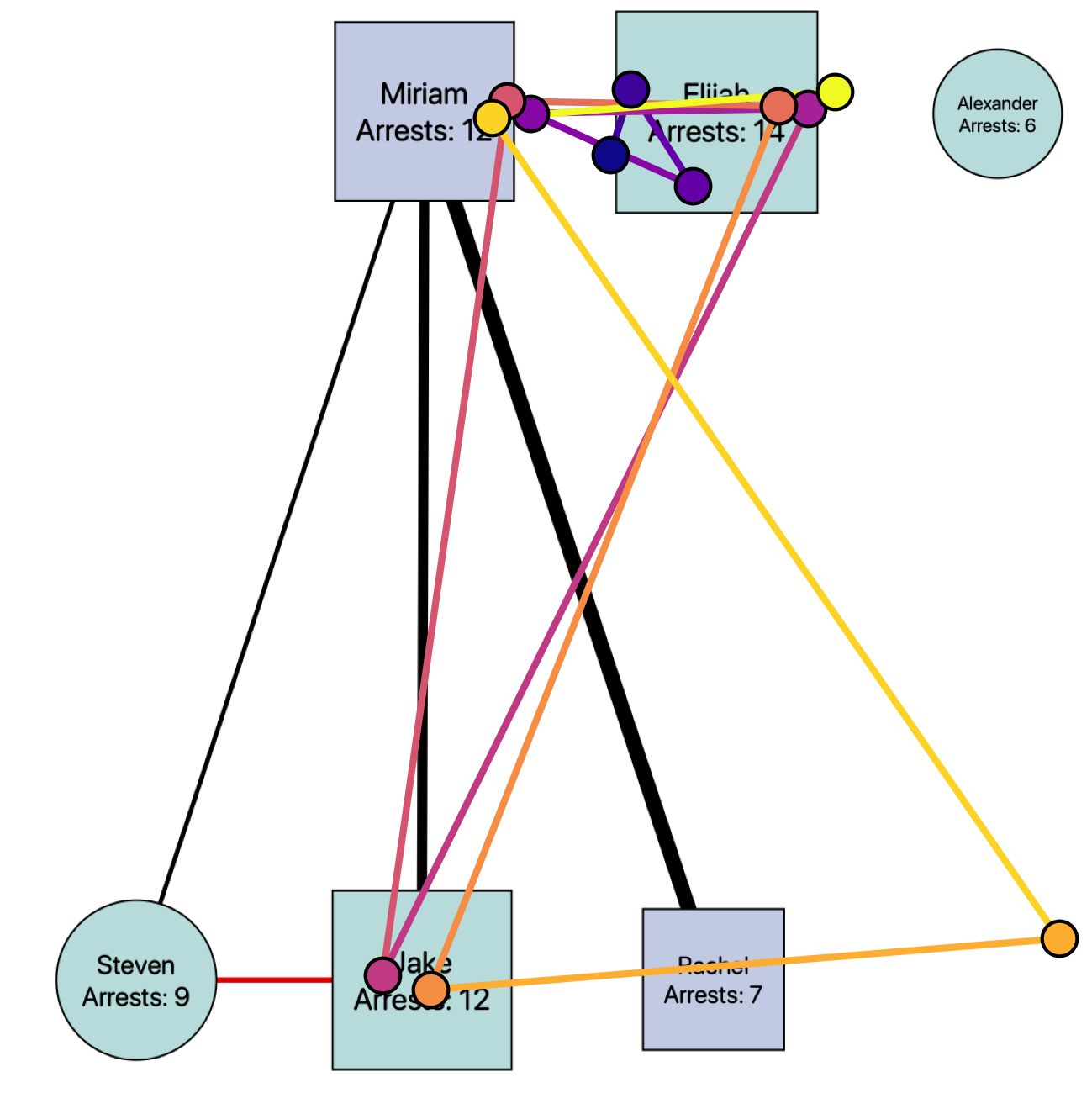}}\hfill
    \subfloat[DeepGaze++\label{fig:scanpath_deepgaze}]{\includegraphics[width=0.225\linewidth]{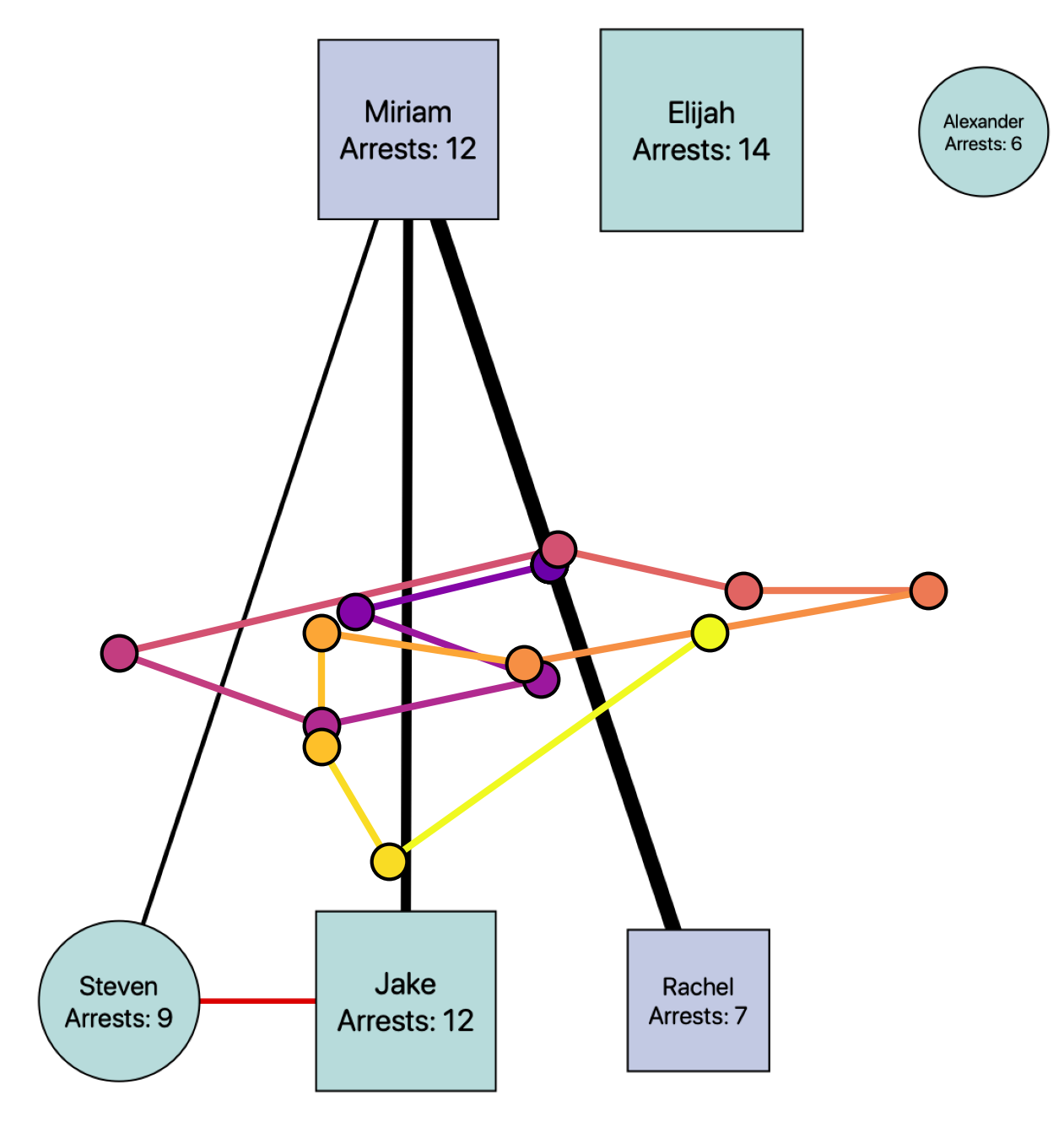}}\hfill
    \subfloat[Gazeformer\label{fig:scanpath_gazeformer}]{\includegraphics[width=0.225\linewidth]{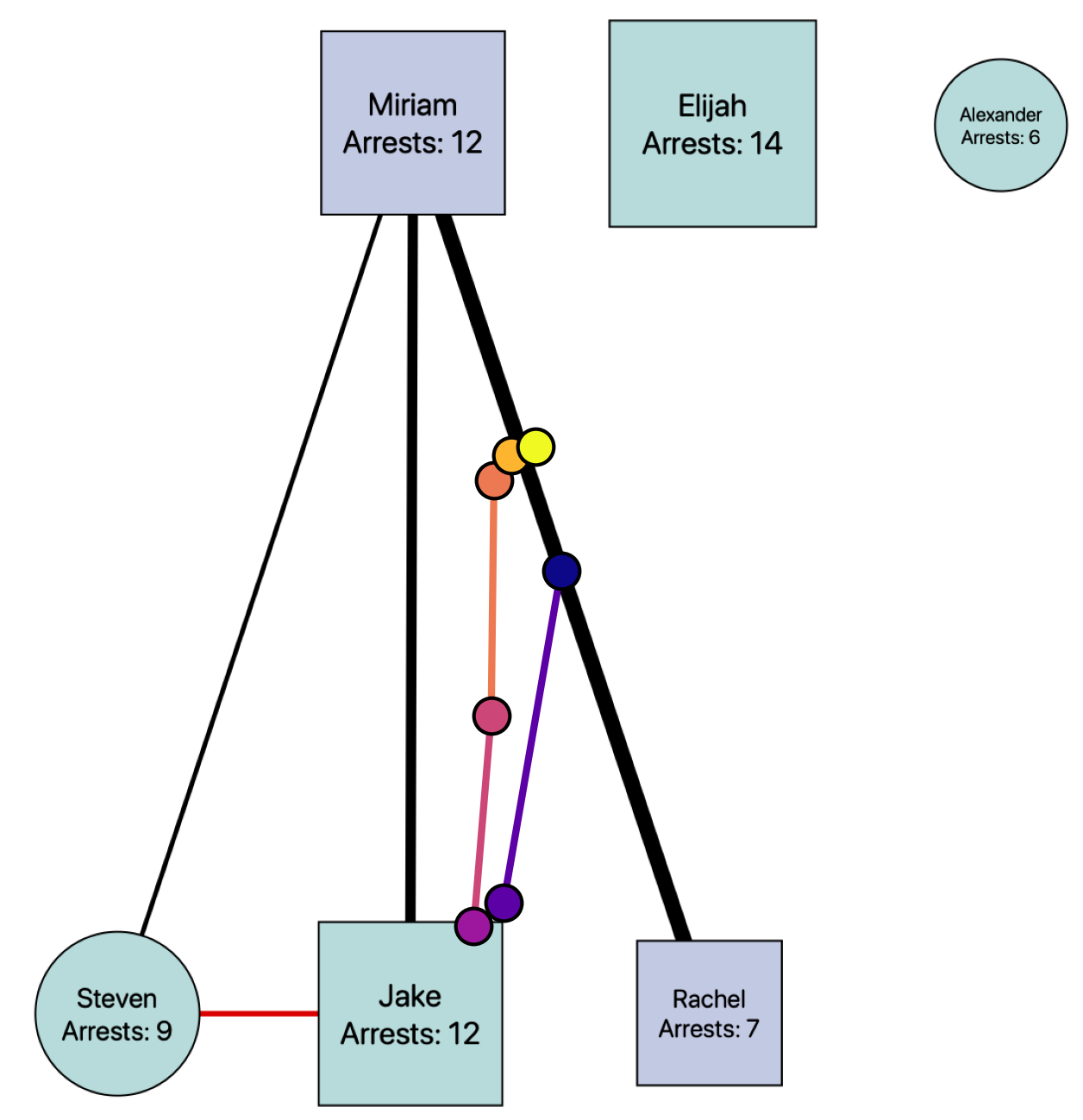}}
 
    \caption{
       Comparison of scanpaths generated by different models on a graph: (a) Actual eye movement sequence (ground truth); (b–d) Scanpaths predicted by UMSS, DeepGaze++, and Gazeformer, respectively. Colored dots represent fixation points, with connecting lines indicating the fixation order. Color intensity represents the temporal order of fixations, with yellow indicating the first point.}
    \label{fig:scanpaths}
\end{figure*}

\subsection{Metrics}
\label{subsec:metrics}
In the case of UMSS and Gazeformer, we generated as many scanpaths per image as there were participants in our study, with the mean evaluation scores representing the averages of all human and predicted scanpath pairs. In contrast, with DeepGaze++, we generated a single scanpath per image. For Gazeformer, we produced seven fixations, as the model is specifically trained to predict this number. In both DeepGaze and UMSS, we generated 12 fixations, reflecting the average fixation count of 12.5 per image in our dataset. We used well-known scanpath prediction metrics used in~\cite{emami_impact_2024}.
\begin{itemize} 
     \item \textbf{Dynamic Time Warping (DTW)}: Introduced by ~\citet{berndt1994using}, DTW compares time series of varying lengths by calculating a distance matrix and finding the optimal path that adheres to boundary, continuity, and monotonicity constraints. The solution is the minimum-cost path from the matrix's start to its endpoint, matching two scanpaths iteratively while including key features~\cite{muller2007dynamic, wang2023improved}. 
    
    \item \textbf{Eyenalysis}: This technique involves double mapping of fixations between two scanpaths to reduce positional variability~\cite{mathot2012simple}. Like DTW, it may assign multiple points from one scanpath to a single point in the other, identifying the closest fixation points and measuring the average distance between corresponding pairs. 
    
    \item \textbf{Determinism:} This metric evaluates diagonal alignments among cross-recurrent points, representing shared fixation trajectories~\cite{fahimi2021metrics}. With a minimum diagonal length of $l_{\min} = 2$, it measures the percentage of recurrent fixation points in sub-scanpaths, enhancing the original unweighted metric by evaluating congruence in fixation sequences. 
    
    \item \textbf{Laminarity:} Laminarity measures the percentage of fixation points in sub-scanpaths where all corresponding pairs are recurrent, sharing the same fixation point from one scanpath~\cite{anderson2015comparison, fahimi2021metrics}. It indicates the tendency of scanpath fixations to cluster in specific locations.
\end{itemize}

\section{Results}
\label{sec:experiments}

The results for the three models, namely UMSS, DeepGaze++, and Gazeformer, are reported in \Cref{tab:results_umss}, \Cref{tab:results_deepgaze}, and \Cref{tab:results_gazeformer}, respectively. The best value for each metric appears in bold, while the second-best is underlined. An example of scanpaths for each model is shown in \Cref{fig:scanpaths}.  To ensure a fair comparison, we present the results of UMSS and DeepGaze++ predicting seven fixations in the Appendix \ref{app_results}. 

 DeepGaze++ achieves the highest quantitative performance, though a visual inspection suggests that UMSS generates more plausible scanpaths (\Cref{fig:scanpaths}). UMSS scanpaths contain more fixations on nodes, aligning with findings from \citet{matzen_data_2018}, who observed that models excelling at natural scenes tend to underperform on data visualisations, and vice versa. Specialised \gls{infovis} models tend to focus more on high-information areas \cite{matzen_data_2018}. This may explain why Gazeformer exhibits the weakest performance in both quantitative metrics and scanpath plausibility (\Cref{fig:scanpaths}), likely due to its training, which lacks \gls{infovis} data, and its limitation of predicting only seven fixations.

In terms of graph complexity, UMSS and DeepGaze++ perform better with six-node graphs and hard-complexity questions, whereas Gazeformer achieves the best results with six-node graphs and easy questions. In three-node graphs, question complexity has less influence on performance. Notably, Gazeformer, the only model that explicitly encodes the question, achieves relatively strong results despite its disadvantages, suggesting that question incorporation plays a role in scanpath prediction.

The evaluated metrics are sensitive to both the number of nodes in the graph and question complexity. Performance improves with six-node stimuli, even when all models predict only seven fixations (Appendix \ref{app_results}). This likely occurs because three-node graphs contain less information than the datasets used to train these models.

Moreover, we find that scanpath predictions are significantly influenced by the training datasets of each model, as well as model-specific parameters such as the number of fixations and internal configuration settings. As \citet{emami_impact_2024} highlight, certain parameters heavily impact predictions and are not easily optimised. For example, as shown in Appendix \ref{app:results_umss} and Appendix \ref{app:results_deepgaze}, both UMSS and DeepGaze++ exhibit metric variations, particularly in DTW, depending on the number of predicted fixations. Additionally, to generate plausible scanpaths with UMSS, we had to adapt the algorithm to our specific visualization type (\autoref{subsec:umss}).

After analysing these three state-of-the-art models---two selected based on their training datasets and one for its ability to incorporate questions---we concluded that using a pretrained scanpath generation model requires careful parameter optimization and potentially fine-tuning to be effective on a new dataset.

\section{Limitations and future work}

When evaluating the models, we did not perform fine-tuning or an exhaustive search for the best parameters. We replicated training or used pretrained checkpoints when available, although with some modifications to adapt them to our data, especially in the case of UMSS.

As future work, it would be interesting to explore how the performance of these models improves when fine-tuned with this dataset. In particular, Gazeformer could be further optimized by training it to predict more than seven fixations and by adapting it specifically for this task. For instance, the model could benefit from fine-tuning the ResNet module used for feature extraction, as it is not optimized for \gls{infovis} images, which differ significantly from the images it was originally trained on. Also, investigate models such as Chartist \cite{shi_chartist_2025} to assess their generalization capabilities on our task and visualization type, different from their training data.

As a complementary analysis, it would be interesting to examine whether user response accuracy affects model performance. Specifically, we could study whether the models predict scanpaths more effectively in trials where the question was answered correctly versus incorrectly.

\section*{Ethical statement}

We obtained informed consent from the participants, ensuring the anonymization of their data. Consequently, the \gls{et} datasets collected in this study provide anonymous records in compliance with ethical board approvals, containing no personal information. This approach guarantees the privacy and anonymity of the collected data. To address potential privacy concerns associated with \gls{et} data collection and processing, the datasets presented are unlinkable to individual participants, thus preventing any possible misuse.

\begin{acks}
This research is supported by Horizon Europe's European Innovation Council through the Pathfinder program (SYMBIOTIK project, grant 101071147) and by the Industrial Doctorate Plan of the Department of Research and Universities of the Generalitat de Catalunya, under Grant AGAUR 2023 DI060.
\end{acks}
\bibliographystyle{ACM-Reference-Format}
\bibliography{bib/refs_eurovis, bib/bibliography}


\appendix
\newpage
\section*{Appendix}

\section{Additional results}
\label{app_results}

\subsection{UMSS}
\label{app:results_umss}
In \autoref{tab_app:results_UMSS_7}, we present the results of UMSS predicting seven fixations. Additionally, we filter the organic data to include only the first seven fixations and we recompute metrics in \autoref{tab_app:results_UMSS_7_filter}. We do the same process for 12 fixations in \autoref{tab_app:results_UMSS_12} and \autoref{tab_app:results_UMSS_12_filter}. The difference in metrics according to the question and the number of nodes is consistent, predicting a different number of fixations. DTW is the metric most affected by both the number of fixations and filtering organic data to the same number of fixations as in the predicted sequence of scanpaths. DTW achieves better values when the number of fixations is similar between the two compared sequences.

\begin{table*}[!h]
\centering
\begin{tabular}{llcccc}
\toprule
Question & Graph & \textbf{DTW ($\downarrow$)} & \textbf{Eyenalysis ($\downarrow$)} & \textbf{Determinism ($\uparrow$)} & \textbf{Laminarity ($\uparrow$)} \\
\midrule
hard & 6 nodes & 3.7558 $\pm$ 0.3765 & \textbf{0.0387 $\pm$ 0.0097} & \textbf{0.5465 $\pm$ 0.6692} & \textbf{7.3626 $\pm$ 3.7459} \\
easy & 6 nodes & \underline{3.4907 $\pm$ 0.5743} & \underline{0.0432 $\pm$ 0.0116} & \underline{0.3046 $\pm$ 0.3219} & \underline{5.1789 $\pm$ 3.2532} \\
hard & 3 nodes & 3.5266 $\pm$ 0.4808 & 0.0475 $\pm$ 0.0117 & 0.1962 $\pm$ 0.2947 & 4.7403 $\pm$ 3.2299 \\
easy & 3 nodes & \textbf{3.0397 $\pm$ 0.4084} & 0.0542 $\pm$ 0.0131 & 0.2104 $\pm$ 0.2948 & 4.8477 $\pm$ 3.8996 \\
\bottomrule
\end{tabular}

\caption{Metrics for the UMSS model (mean $\&$ standard deviation) predicting 7 fixations, evaluated by graph complexity and question difficulty.}
\label{tab_app:results_UMSS_7}
\end{table*}
\begin{table*}[!h]
\centering
\begin{tabular}{llcccc}
\toprule
Question & Graph & \textbf{DTW ($\downarrow$)} & \textbf{Eyenalysis ($\downarrow$)} & \textbf{Determinism ($\uparrow$)} & \textbf{Laminarity ($\uparrow$)} \\
\midrule
hard & 6 nodes & \textbf{2.3068 $\pm$ 0.2620} & \textbf{0.0483 $\pm$ 0.0100} & \textbf{0.5465 $\pm$ 0.6692} & \textbf{7.3626 $\pm$ 3.7459} \\
easy & 6 nodes & \underline{2.4121 $\pm$ 0.3016} & \underline{0.0548 $\pm$ 0.0140} & \underline{0.3046 $\pm$ 0.3219} & \underline{5.1789 $\pm$ 3.2532} \\
hard & 3 nodes & 2.6027 $\pm$ 0.1689 & 0.0586 $\pm$ 0.0068 & 0.1962 $\pm$ 0.2947 & 4.7403 $\pm$ 3.2299 \\
easy & 3 nodes & 2.5613 $\pm$ 0.1503 & 0.0620 $\pm$ 0.0096 & 0.2104 $\pm$ 0.2948 & 4.8477 $\pm$ 3.8996 \\
\bottomrule
\end{tabular}

\caption{Metrics for the UMSS model (mean $\&$ standard deviation) predicting 7 fixations, evaluated by graph complexity and question difficulty, filtering organic data based on the first 7 fixations.}
\label{tab_app:results_UMSS_7_filter}
\end{table*}

\begin{table*}[!h]
\centering
\begin{tabular}{llcccc}
\toprule
Question & Graph & \textbf{DTW ($\downarrow$)} & \textbf{Eyenalysis ($\downarrow$)} & \textbf{Determinism ($\uparrow$)} & \textbf{Laminarity ($\uparrow$)} \\
\midrule
hard & 6 nodes & 4.3676 $\pm$ 0.4010 & \textbf{0.0325 $\pm$ 0.0071} & \textbf{1.6162 $\pm$ 1.4088} & \textbf{13.3448 $\pm$ 5.9174} \\
easy & 6 nodes & \underline{4.3009 $\pm$ 0.5005} & \underline{0.0352 $\pm$ 0.0100} & \underline{1.0073 $\pm$ 0.9065} & \underline{8.7668 $\pm$ 4.0127} \\
hard & 3 nodes & 4.3414 $\pm$ 0.4041 & 0.0373 $\pm$ 0.0077 & 0.8511 $\pm$ 0.7929 & 8.5293 $\pm$ 4.7948 \\
easy & 3 nodes & \textbf{4.1614 $\pm$ 0.3345} & 0.0454 $\pm$ 0.0143 & 0.6084 $\pm$ 0.7561 & 7.0486 $\pm$ 4.8389 \\
\bottomrule
\end{tabular}

\caption{Metrics for the UMSS model (mean $\&$ standard deviation) predicting 12 fixations, evaluated by graph complexity and question difficulty.}
\label{tab_app:results_UMSS_12}
\end{table*}
\begin{table*}[!h]
\centering
\begin{tabular}{llcccc}
\toprule
Question & Graph & \textbf{DTW ($\downarrow$)} & \textbf{Eyenalysis ($\downarrow$)} & \textbf{Determinism ($\uparrow$)} & \textbf{Laminarity ($\uparrow$)} \\
\midrule
hard & 6 nodes & \textbf{3.8517 $\pm$ 0.3717} & \textbf{0.0366 $\pm$ 0.0074} & \textbf{1.6162 $\pm$ 1.4088} & \textbf{13.3448 $\pm$ 5.9174} \\
easy & 6 nodes & \underline{3.9480 $\pm$ 0.4351} & \underline{0.0391 $\pm$ 0.0095} & \underline{1.0073 $\pm$ 0.9065} & \underline{8.7668 $\pm$ 4.0127} \\
hard & 3 nodes & 4.0980 $\pm$ 0.3104 & 0.0399 $\pm$ 0.0067 & 0.8511 $\pm$ 0.7929 & 8.5293 $\pm$ 4.7948 \\
easy & 3 nodes & 4.0598 $\pm$ 0.3011 & 0.0467 $\pm$ 0.0139 & 0.6084 $\pm$ 0.7561 & 7.0486 $\pm$ 4.8389 \\
\bottomrule
\end{tabular}

\caption{Metrics for the UMSS model (mean $\&$ standard deviation) predicting 12 fixations, evaluated by graph complexity and question difficulty, filtering organic data based on the first 12 fixations.}
\label{tab_app:results_UMSS_12_filter}
\end{table*}

\subsection{DeepGaze++}
\label{app:results_deepgaze}
In \autoref{tab_app:results_deepgaze_7}, we present the results of DeepGaze++ predicting seven fixations. Additionally, we filter the organic data to include only the first seven fixations and we recompute metrics in \autoref{tab_app:results_deepgaze_7_filter}. We do the same process for 12 fixations in \autoref{tab_app:results_deepgaze_12} and \autoref{tab_app:results_deepgaze_12_filter}. Similarly as explained in \autoref{app:results_umss}, the most affected metric by filtering is DTW.

\begin{table*}[h]
\centering
\begin{tabular}{llcccc}
\toprule
Question & Graph & \textbf{DTW ($\downarrow$)} & \textbf{Eyenalysis ($\downarrow$)} & \textbf{Determinism ($\uparrow$)} & \textbf{Laminarity ($\uparrow$)} \\
\midrule
hard & 6 nodes & 4.4279 $\pm$0.7542 & \textbf{0.0296 $\pm$0.0086} & \textbf{5.2506 $\pm$2.7208} & \textbf{21.6753 $\pm$7.7796} \\
easy & 6 nodes & \underline{3.6398 $\pm$0.9210} & 0.0355 $\pm$0.0101 & \underline{4.5245 $\pm$3.0246} & \underline{19.7765 $\pm$6.2049} \\
hard & 3 nodes & 3.7249 $\pm$0.8538 & 0.0392 $\pm$0.0112 & 3.2529 $\pm$2.6520 & 14.1096 $\pm$4.8151 \\
easy & 3 nodes & \textbf{2.8477 $\pm$0.5451} & \underline{0.0350 $\pm$0.0100} & 3.4354 $\pm$2.1081 & 17.8470 $\pm$4.9186 \\
\bottomrule
\end{tabular}

\caption{Metrics for the DeepGaze++ model (mean $\&$ standard deviation) predicting 7 fixations, evaluated by graph complexity and question difficulty.}

\label{tab_app:results_deepgaze_7}
\end{table*}
\begin{table*}[h]
\centering
\begin{tabular}{llcccc}
\toprule
Question & Graph & \textbf{DTW ($\downarrow$)} & \textbf{Eyenalysis ($\downarrow$)} & \textbf{Determinism ($\uparrow$)} & \textbf{Laminarity ($\uparrow$)} \\
\midrule
hard & 6 nodes & \textbf{2.1105 $\pm$0.2562} & \textbf{0.0264 $\pm$0.0077} & \textbf{4.8746 $\pm$3.5018} & \textbf{20.4399 $\pm$6.7628} \\
easy & 6 nodes & 2.3628 $\pm$0.3176 & 0.0350 $\pm$0.0096 & \underline{4.4420 $\pm$3.0489} & \underline{19.2532 $\pm$6.1302} \\
hard & 3 nodes & 2.4614 $\pm$0.3535 & 0.0369 $\pm$0.0110 & 3.2124 $\pm$2.5604 & 14.1334 $\pm$5.3275 \\
easy & 3 nodes & \underline{2.2919 $\pm$0.3018} & \underline{0.0335 $\pm$0.0098} & 3.6108 $\pm$2.2370 & 18.1738 $\pm$5.3647 \\
\bottomrule
\end{tabular}

\caption{Metrics for the DeepGaze++ model (mean $\&$ standard deviation) predicting 7 fixations, evaluated by graph complexity and question difficulty, filtering organic data based on the first 7 fixations.}
\label{tab_app:results_deepgaze_7_filter}
\end{table*}

\begin{table*}[h]
\centering
\begin{tabular}{llcccc}
\toprule
Question & Graph & \textbf{DTW ($\downarrow$)} & \textbf{Eyenalysis ($\downarrow$)} & \textbf{Determinism ($\uparrow$)} & \textbf{Laminarity ($\uparrow$)} \\
\midrule
hard & 6 nodes & 4.7764 $\pm$0.7233 & \textbf{0.0253 $\pm$0.0078} & \textbf{5.7138 $\pm$2.8525} & \textbf{22.3863 $\pm$7.9269} \\
easy & 6 nodes & \underline{4.3233 $\pm$0.8756} & \underline{0.0315 $\pm$0.0098} & \underline{4.7295 $\pm$2.8249} & \underline{19.5152 $\pm$6.1463} \\
hard & 3 nodes & 4.4060 $\pm$0.6973 & 0.0348 $\pm$0.0085 & 3.8870 $\pm$2.2667 & 13.9008 $\pm$4.7745 \\
easy & 3 nodes & \textbf{3.6790 $\pm$0.4919} & 0.0327 $\pm$0.0084 & 3.6320 $\pm$2.3660 & 17.8022 $\pm$4.7747 \\
\bottomrule
\end{tabular}

\caption{Metrics for the DeepGaze++ model (mean $\&$ standard deviation) predicting 12 fixations, evaluated by graph complexity and question difficulty.}
\label{tab_app:results_deepgaze_12}
\end{table*}
\begin{table*}[h]
\centering
\begin{tabular}{llcccc}
\toprule
Question & Graph & \textbf{DTW ($\downarrow$)} & \textbf{Eyenalysis ($\downarrow$)} & \textbf{Determinism ($\uparrow$)} & \textbf{Laminarity ($\uparrow$)} \\
\midrule
hard & 6 nodes & \underline{3.3961 $\pm$0.4109} & \textbf{0.0250 $\pm$0.0069} & \textbf{5.4042 $\pm$2.9753} & \textbf{21.8460 $\pm$7.6795} \\
easy & 6 nodes & 3.6931 $\pm$0.5609 & \underline{0.0321 $\pm$0.0099} & \underline{4.5699 $\pm$2.9623} & \underline{19.6052 $\pm$6.3182} \\
hard & 3 nodes & 3.8254 $\pm$0.4870 & 0.0342 $\pm$0.0090 & 3.3734 $\pm$2.2944 & 13.9846 $\pm$4.8498 \\
easy & 3 nodes & \textbf{3.4742 $\pm$0.4003} & 0.0324 $\pm$0.0084 & 3.3515 $\pm$2.0423 & 17.7992 $\pm$4.9506 \\
\bottomrule
\end{tabular}

\caption{Metrics for the DeepGaze++ model (mean $\&$ standard deviation) predicting 12 fixations, evaluated by graph complexity and question difficulty, filtering organic data based on the first 12 fixations.}
\label{tab_app:results_deepgaze_12_filter}
\end{table*}

\subsection{Gazeformer}

In \autoref{tab_app:results_gazeformer_7}, we present the results of Gazeformer predicting seven fixations. Additionally, we filter the organic data to include only the first seven fixations and we recompute metrics in \autoref{tab_app:results_gazeformer_7_filter}.

\begin{table*}[h]
\centering
\begin{tabular}{llcccc}
\toprule
Question & Graph & \textbf{DTW ($\downarrow$)} & \textbf{Eyenalysis ($\downarrow$)} & \textbf{Determinism ($\uparrow$)} & \textbf{Laminarity ($\uparrow$)} \\
\midrule
hard & 6 nodes & 4.9047 $\pm$0.9595 & \underline{0.0392 $\pm$0.0121} & 0.3598 $\pm$0.9258 & \underline{6.4004 $\pm$4.3728} \\
easy & 6 nodes & \underline{3.3858 $\pm$1.2919} & \textbf{0.0379 $\pm$0.0161} & \textbf{1.2398 $\pm$2.2932} & \textbf{7.7372 $\pm$6.3383} \\
hard & 3 nodes & 3.5733 $\pm$1.1852 & 0.0414 $\pm$0.0132 & 0.7707 $\pm$1.8329 & 4.3177 $\pm$3.7336 \\
easy & 3 nodes & \textbf{2.6133 $\pm$0.7640} & 0.0421 $\pm$0.0167 & \underline{0.9666 $\pm$2.3136} & 5.5023 $\pm$5.1348 \\
\bottomrule
\end{tabular}

\caption{Metrics for the Gazeformer model (mean $\&$ standard deviation) predicting 7 fixations, evaluated by graph complexity and question difficulty.}
\label{tab_app:results_gazeformer_7}
\end{table*}
\begin{table*}[h]
\centering
\begin{tabular}{llcccc}
\toprule
Question & Graph & \textbf{DTW ($\downarrow$)} & \textbf{Eyenalysis ($\downarrow$)} & \textbf{Determinism ($\uparrow$)} & \textbf{Laminarity ($\uparrow$)} \\
\midrule
hard & 6 nodes & 2.4586 $\pm$0.5782 & 0.0555 $\pm$0.0223 & 0.3598 $\pm$0.9258 & \underline{6.4004 $\pm$4.3728} \\
easy & 6 nodes & \textbf{2.1863 $\pm$0.7229} & \textbf{0.0490 $\pm$0.0246} & \textbf{1.2398 $\pm$2.2932 }& \textbf{7.7372 $\pm$6.3383} \\
hard & 3 nodes & 2.3663 $\pm$0.5949 & 0.0519 $\pm$0.0213 & 0.7707 $\pm$1.8329 & 4.3177 $\pm$3.7336\\
easy & 3 nodes & \underline{2.1944 $\pm$0.5489} & \underline{0.0504 $\pm$0.0221} & \underline{0.9666 $\pm$2.3136} & 5.5023 $\pm$5.1348 \\
\bottomrule
\end{tabular}
\caption{Metrics for the Gazeformer model (mean $\&$ standard deviation) predicting 7 fixations, evaluated by graph complexity and question difficulty, filtering organic data based on the first 7 fixations.}

\label{tab_app:results_gazeformer_7_filter}
\end{table*}

\end{document}